\documentclass[10pt]{article}
\usepackage[utf8]{inputenc}
\usepackage[T1]{fontenc}

\usepackage[letterpaper]{geometry}
\usepackage{hicss}
\usepackage{times}
\usepackage[none]{hyphenat}
\usepackage{url}
\usepackage{latexsym}
\usepackage{indentfirst}
\usepackage{graphicx}
\graphicspath{{images/}}
\usepackage{amsmath,amssymb}
\usepackage{booktabs}
\usepackage{array}
\usepackage{float}
\usepackage{caption}
\usepackage{subcaption}
\usepackage{pgfplots}
\pgfplotsset{compat=1.18}
\usepgfplotslibrary{groupplots}
\usepackage[
  style=apa,
  natbib=true,
  backend=biber,
  doi=true,
  url=true,
  isbn=false
]{biblatex}
\DeclareLanguageMapping{american}{american-apa}
\pgfplotsset{
  compact/.style={
    width=\linewidth, height=5.2cm,
    grid=both,
    grid style={line width=.1pt, draw=gray!25},
    major grid style={line width=.2pt, draw=gray!45},
    scaled x ticks=false,
    scaled y ticks=false,
    tick label style={font=\footnotesize,/pgf/number format/fixed,/pgf/number format/precision=3},
    label style={font=\footnotesize},
    title style={font=\footnotesize\bfseries, yshift=-2pt},
    legend style={font=\scriptsize, legend cell align=left, draw=none,
                  fill=white, inner sep=1pt, row sep=-1pt},
    every axis plot/.append style={line width=0.9pt},
  },
}

\title{Beyond the Illusion of Power: \\Calibrating Quasi-Experiments in Observational IS}

\author{Spandan Ghose Chowdhury \\
  Georgia Institute of Technology, GA, USA \\
  {\underline{spandan\_gc@gatech.edu}} \\}

\date{}

\begin{document}
\maketitle

\begin{abstract}
Information systems (IS) researchers increasingly use quasi-experimental methods such as difference-in-differences (DiD) and instrumental variables (IV) to recover causal effects from observational panel data. Power calculations that justify these designs assume i.i.d.\ errors, but the deeper problem is what even a cluster-robust calculator cannot see. We report a Monte Carlo study over 9,837 parameter conditions ($\approx$9.8 million datasets) and decompose
the planned-versus-achieved power gap. The serial-correlation component is recoverable by an AR(1)-aware calculator when $\rho$ is known, and partially when $\rho$ must be estimated from short pre-periods; but panel attrition, staggered-adoption bias, and parallel-trends pretesting are captured by no closed-form formula; exogenous attrition alone costs $\approx$8–11 percentage points at the few-hundred-to-thousand sample sizes IS studies use. Treatment-correlated, outcome-dependent attrition instead induces bias, not just power loss. For IV, holding first-stage $F$ fixed, larger $N$ neither raises power nor curbs exclusion bias---though with a \emph{fixed instrument} more data does sharpen the first stage---so identification rests on instrument strength, not sample size. We provide a unified
DiD+IV benchmark calibrated to IS effect sizes, a companion lookup tool, and design guidance for observational causal work in IS: \url{https://github.com/spandan1305/is-power-lookup}
\end{abstract}

\subsubsection*{Keywords:}

Power analysis, difference-in-differences, instrumental variables, Monte Carlo
simulation, research methodology.

\section{Introduction}

Causal designs now occupy a central position in Information Systems (IS)
empirical research \citep{tafti2025}, particularly through
difference-in-differences (DiD) and, where endogeneity concerns are stronger,
instrumental variables (IV). Yet ex ante design justification often remains
anchored in power calculations derived from i.i.d.\ settings---poorly aligned
with IS panels, where outcomes are serially correlated, attrition unbalances
the panel, treatment timing is staggered, and effects are heterogeneous.

Our aim is not to re-establish that i.i.d.\ calculators are optimistic---that
much is well known. We ask instead how much of the planned-versus-achieved
power gap survives a competent, cluster-robust, AR(1)-aware calculator, and
which frictions remain invisible to any closed-form formula. First, once the
recoverable serial-correlation component is removed, how much power loss
persists from panel attrition and staggered timing? Second, what is the
inferential consequence of conditioning DiD analysis on pretrend-test
non-rejection \citep{roth2022}? Third, how do weak-instrument and
exclusion-restriction failures in IV differ as sample size increases?

Several methodological studies provide partial answers. \citet{bertrand2004}
document serial-correlation distortions in DiD inference; \citet{roth2022}
formalizes the conditional-inference problem in pretrend testing; and
\citet{goodmanbacon2021}, \citet{dechaisemartin2020}, and \citet{sun2021}
characterize TWFE bias under staggered timing. What is missing is a
\emph{joint} quantification that treats panel frictions, estimator choice,
conditional inference from pretesting, and IV identification failures as
components of a single design problem, evaluated over an IS-calibrated
parameter space with consistent metrics.

The individual mechanisms we study are established in econometrics---serial-correlation distortion \citep{bertrand2004}, pretest selection \citep{roth2022}, TWFE bias under staggered timing \citep{goodmanbacon2021}, and the weak-versus-exclusion IV trade-off. Our contribution is to quantify their \emph{joint} magnitude in IS-typical panels and package that evidence for practice. First, we provide a unified Monte Carlo benchmark covering both DiD and IV under one IS-realistic data-generating process (DGP), reporting power, bias, root-mean-squared error, and confidence interval coverage over 9{,}837 parameter conditions. Second, we calibrate the effect-size grid to documented IS effect magnitudes (Table~\ref{tab:is-anchors}), validate the simulation against canonical open datasets, and release a companion lookup tool over the full grid; the tool matches categorical inputs exactly, \emph{snaps} numeric inputs to the nearest simulated grid value, and reports the matched design so users can judge how far the returned cell sits from their target. Third, we document an asymmetry with direct bearing on IV practice: weak-instrument bias attenuates with $N$, while exclusion bias is not cured by sample size, so identification rests on instrument strength rather than $N$. We translate these results into conditional design guidance rather than a universal checklist.

\section{Background}

In contemporary IS empirical work, two-way fixed effects (TWFE) DiD remains the
default framework for policy rollouts, platform interventions, and staged
feature launches. In the canonical model
$Y_{it}=\alpha_i+\lambda_t+\delta D_{it}+\varepsilon_{it}$, identification of
$\delta$ rests on parallel trends, typically assessed with event-study plots
and joint tests on pre-treatment coefficients---diagnostics that are
themselves finite-sample objects and noisy in observational panels.

Once timing is staggered and effects are heterogeneous, TWFE no longer has a
straightforward ATT interpretation: it combines multiple 2x2 comparisons,
including already-treated vs.\ not-yet-treated contrasts that can receive
negative weight and induce bias \citep{goodmanbacon2021,dechaisemartin2020}.
We therefore benchmark TWFE against alternatives that re-center
identification at the group-time level, notably \citet{sun2021} and
\citet{callaway2021}; for a recent systematic comparison see
\citet{ulloaperez2025}.

Pretrend testing introduces a second inferential issue: \citet{roth2022}
shows that conditioning interpretation on jointly insignificant pre-treatment
coefficients induces selection bias, because samples that pass are those in
which sampling noise masks true trend differences. Conditional error rates
can then exceed unconditional ones, especially when pre-periods are short and
noisy --- conditions common in IS panels.

IV designs face a related dual-identification challenge: relevance and
exclusion. In the just-identified case,
$\hat{\beta}_{IV}=(Z'X)^{-1}Z'Y$, relevance is often summarized through
first-stage diagnostics such as \citet{stockyogo2005}, while exclusion
requires that $Z$ affect $Y$ only through $X$. IS instruments are often
geographic, temporal, or peer-exposure based and plausibly vulnerable to
contextual spillovers \citep{semadeni2014}, making first-stage strength and
exclusion plausibility distinct empirical concerns.

Because serially dependent, attrition-thinned panels reduce effective
information relative to nominal row counts \citep{bertrand2004}, our
simulation treats serial dependence, attrition, and clustering as first-order
design inputs rather than implementation details.

Beyond Sun--Abraham and Callaway--Sant'Anna, the DiD literature has expanded
to include imputation-based approaches \citep{borusyak2024}, two-stage DiD
\citep{gardner2022}, panel counterfactual methods rooted in matrix completion
and synthetic-control logic \citep{athey2021,abadie2021}, and causal machine
learning \citep{chernozhukov2018,wagerathey2018,vonzahn2026}. We do not
implement these classes; the present grid is a benchmark against which they
can be compared in follow-up work.

Finally, we extend simulation-based design diagnostics in management and IS
\citep{semadeni2014} by jointly analyzing DiD and IV on one panel grid that
models serial correlation and attrition and quantifies both estimator
performance and inference calibration.

\section{Simulation Design}

\subsection{Data-Generating Process}

The DiD DGP generates
$Y_{it} = \alpha_i + \lambda_t + \delta D_{it} + \varepsilon_{it}$ with
$\varepsilon_{it} = \rho \varepsilon_{i,t-1} + u_{it}$,
$u_{it} \sim \mathcal{N}(0, \sigma^2(1-\rho^2))$,
$\alpha_i \sim \mathcal{N}(0,1)$, and $\lambda_t = 0.5\, t/T$. To span the
design space most frequently encountered in IS panels, we vary nine inputs:
sample size (\texttt{n\_units} $\in \{50, 100, 200, 500, 1000, 5000\}$),
panel length (\texttt{n\_periods} $\in \{6, 10, 16, 24\}$), pre-period share
(\texttt{pre\_frac} $\in \{0.33, 0.50\}$), standardized effect magnitude
(\texttt{effect\_size} $\in \{0.1, 0.2, 0.5\}$ SD), serial persistence
(\texttt{autocorr} $\rho \in \{0, 0.3, 0.5, 0.7\}$), adoption timing
(\texttt{staggered} $\in \{\text{False}, \text{True}\}$), treatment-effect
heterogeneity (\texttt{het\_effects}, with early cohorts receiving $2\delta$
and later cohorts $\delta$), exogenous attrition
(\texttt{churn\_rate} $\in \{0, 0.3\}$), and covariance structure for
inference (\texttt{cluster\_type} $\in \{\text{unit}, \text{two-way}\}$).
After removing infeasible cells, this yields $6{,}912$ DiD conditions. (\texttt{het\_effects} is defined only under staggered adoption, which removes one quarter of the $9{,}216$ nominal cells.) We
additionally run a 45-cell pretrend sub-study and a $2{,}880$-cell IV grid
that varies first-stage strength $F$, endogeneity $\rho_{vu}$, sample size,
instrument count, and a direct $Z \to Y$ channel. Crucially, the grid fixes the \emph{target} first-stage $F$ rather than the first-stage coefficient, so the instrument strength is set to $\pi=\sqrt{F/(N+F)}$, which declines as $N^{-1/2}$; this normalization drives the IV scaling result below.

\subsection{Estimators}

For DiD, we estimate (i) TWFE under unit-clustered or two-way clustered
covariance, (ii) the interaction-weighted event-study estimator of
\citet{sun2021}, and (iii) cohort-specific ATT aggregation following
\citet{callaway2021}. Pretrend diagnostics use a joint Wald test over all
pre-treatment event-time coefficients in a two-way FE event-study
regression. For IV, we estimate just-identified and over-identified
(three-instrument) 2SLS with heteroskedasticity-robust standard errors.

To calibrate the standardized effect-size grid to effects plausibly encountered in IS research, Table~\ref{tab:is-anchors} maps each range to a representative study type. The anchors come from a structured (though not systematic) review of published IS quasi-experiments and large-scale digital experiments, combined with domain judgment: massive online field experiments report standardized effects at the low end \citep{aralwalker2011}, platform- and policy-scale interventions on short persistent panels fall in the middle ranges \citep{aziz2023, huang2017, seamanszhu2014}, and effects above $0.5$~SD are rare in credible causal designs \citep{gelmancarlin2014}. Researchers with better priors should query the lookup tool at those values instead.

\begin{table}[t]
\centering
\small
\begin{tabular}{@{}lp{5.2cm}@{}}
\toprule
\textbf{Effect (SD)} & \textbf{IS example} \\
\midrule
0.05--0.10 & UX A/B test on conversion (button, copy) \\
0.10--0.20 & Recommender change on click-through \\
0.20--0.35 & Platform policy change (pricing, matching) \\
0.35--0.50 & Major ERP/system implementation on firm productivity \\
$>0.50$      & Rarely observed in credible causal designs \\
\bottomrule
\end{tabular}
\caption{IS effect-size anchors: illustrative calibration points from a
structured review plus domain judgment (see text), not meta-analytic
estimates.}
\label{tab:is-anchors}
\end{table}

\subsection{Performance Metrics}

For each design cell, we run $N_{\text{sim}} = 1{,}000$ independent Monte Carlo
replications (500 in the dedicated pretrend exercise) and report empirical
power ($\hat{P}(\text{reject } H_0:\beta=0)$), bias
($\mathbb{E}[\hat\beta-\beta]$), RMSE, and 95\% confidence-interval coverage.
In the pretrend module we additionally compute the unconditional versus conditional rejection gap emphasized by \citet{roth2022}. The full experiment
covers $9{,}837$ parameter conditions ($6{,}912$ DiD $+$ $45$ pretrend $+$
$2{,}880$ IV cells) and approximately 9.8 million simulated datasets. Three supplementary sub-studies---a 48-cell attrition grid (Section~\ref{sec:attrition}), a 500-pilot-per-cell estimated-$\rho$ analysis (Table~\ref{tab:rho-sensitivity}), and a ten-cell reference-estimator validation (Section~\ref{sec:scaling})---lie outside this count and are described where used.

\subsection{Validation}

Before launching the full grid, we validate implementation fidelity using
five unit tests summarized in Table~\ref{tab:validation}. All pass within
their pre-registered tolerances.

\begin{table}[t]
\centering
\small
\begin{tabular}{@{}llll@{}}
\toprule
Test & Target & Achieved & Tol. \\
\midrule
High-power recovery       & $\ge 0.90$    & 0.94   & $\pm 0.03$ \\
Type~I ($\beta=0$)  & $\approx 0.05$ & 0.047 & $\pm 0.01$ \\
95\% CI coverage          & $\approx 0.95$ & 0.955 & $\pm 0.01$ \\
Churn symmetry & diff $< 0.02$ & 0.008 & --- \\
AR(1) $\rho$ recovery     & $|\hat\rho - \rho_0| < 0.05$ & 0.023 & --- \\
\bottomrule
\end{tabular}
\caption{Pre-flight validation results against five implementation-fidelity
targets. Each test uses 1{,}000 replications.}
\label{tab:validation}
\end{table}

\section{DiD Results}
\label{sec:did-results}

\subsection{Sample-Size Scaling Under Serial Correlation and Staggered Adoption}
\label{sec:scaling}

\begin{figure*}[t]
\centering
\begin{subfigure}[t]{0.32\linewidth}
\centering
\begin{tikzpicture}
\begin{axis}[compact, scale only axis, width=3.2cm, height=3.0cm,
  xmode=log, log ticks with fixed point,
  xlabel={$N$ (log scale)}, ylabel={TWFE power},
  xtick={50,200,1000,5000},
  xticklabel style={font=\tiny},
  yticklabel style={font=\tiny},
  label style={font=\scriptsize},
  ymin=0, ymax=1.05,
  xlabel style={yshift=2pt},
  ylabel style={yshift=-4pt},
  legend style={at={(0.5,-0.38)},anchor=north,font=\tiny,
    column sep=2pt, /tikz/every even column/.append style={column sep=4pt}},
  legend columns=2]
\addplot[color=blue!70!black, mark=*]     coordinates {(50,0.260)(100,0.428)(200,0.659)(500,0.917)(1000,0.988)(5000,1.000)}; \addlegendentry{$\rho=0.0$}
\addplot[color=teal,           mark=square*] coordinates {(50,0.167)(100,0.286)(200,0.508)(500,0.834)(1000,0.969)(5000,1.000)}; \addlegendentry{$\rho=0.3$}
\addplot[color=orange!80!black,mark=triangle*] coordinates {(50,0.135)(100,0.226)(200,0.401)(500,0.762)(1000,0.956)(5000,1.000)}; \addlegendentry{$\rho=0.5$}
\addplot[color=red!80!black,   mark=diamond*]  coordinates {(50,0.117)(100,0.205)(200,0.340)(500,0.699)(1000,0.946)(5000,1.000)}; \addlegendentry{$\rho=0.7$}
\addplot[dashed, gray] coordinates {(50,0.8) (5000,0.8)};
\end{axis}
\end{tikzpicture}
\caption{TWFE power vs.\ $N$, $\rho$.}
\label{fig:fig1}
\end{subfigure}\hfill
\begin{subfigure}[t]{0.32\linewidth}
\centering
\begin{tikzpicture}
\begin{axis}[compact, scale only axis, width=3.2cm, height=3.0cm,
  xmode=log, log ticks with fixed point,
  xlabel={$N$ (log scale)}, ylabel={$|\hat\delta-\delta|$ (SD)},
  xtick={50,200,1000,5000},
  xticklabel style={font=\tiny},
  yticklabel style={font=\tiny},
  label style={font=\scriptsize},
  ymin=0.025, ymax=0.065,
  xlabel style={yshift=2pt},
  ylabel style={yshift=-4pt},
  legend style={at={(0.5,-0.38)},anchor=north,font=\tiny,
    column sep=2pt, /tikz/every even column/.append style={column sep=4pt}},
  legend columns=2]
\addplot[color=red!80!black, mark=*]        coordinates {(50,0.0401)(100,0.0424)(200,0.0419)(500,0.0413)(1000,0.0417)(5000,0.0410)}; \addlegendentry{TWFE}
\addplot[color=orange!85!black, mark=square*] coordinates {(50,0.0535)(100,0.0544)(200,0.0552)(500,0.0539)(1000,0.0546)(5000,0.0536)}; \addlegendentry{Sun \& Abr.}
\addplot[color=teal, mark=triangle*]        coordinates {(50,0.0348)(100,0.0336)(200,0.0337)(500,0.0326)(1000,0.0323)(5000,0.0321)}; \addlegendentry{CS}
\end{axis}
\end{tikzpicture}
\caption{Bias (staggered, het.\ effects).}
\label{fig:fig2}
\end{subfigure}\hfill
\begin{subfigure}[t]{0.32\linewidth}
\centering
\begin{tikzpicture}
\begin{axis}[compact, scale only axis, width=3.2cm, height=3.0cm,
  xmode=log, log ticks with fixed point,
  xlabel={$N$ (log scale)}, ylabel={TWFE power},
  xtick={50,200,1000,5000},
  xticklabel style={font=\tiny},
  yticklabel style={font=\tiny},
  label style={font=\scriptsize},
  ymin=0, ymax=1.05,
  xlabel style={yshift=2pt},
  ylabel style={yshift=-4pt},
  legend style={at={(0.5,-0.38)},anchor=north,font=\tiny,
    column sep=2pt, /tikz/every even column/.append style={column sep=4pt}},
  legend columns=1]
\addplot[color=blue!70!black, mark=*] coordinates {(50,0.167)(100,0.280)(200,0.491)(500,0.822)(1000,0.963)(5000,1.000)}; \addlegendentry{churn $=0$}
\addplot[color=red!80!black, mark=square*, dashed] coordinates {(50,0.139)(100,0.230)(200,0.411)(500,0.742)(1000,0.931)(5000,1.000)}; \addlegendentry{churn $=0.3$}
\addplot[dashed, gray] coordinates {(50,0.8) (5000,0.8)};
\end{axis}
\end{tikzpicture}
\caption{TWFE power loss from attrition.}
\label{fig:fig7}
\end{subfigure}
\caption{DiD precision under panel frictions. (a) TWFE power decays steeply
with residual autocorrelation (effect size~$=0.2$~SD, non-staggered, 50\%
pre-period). (b) Deviation from the base effect $\delta$ (not the estimand) under staggered, heterogeneous effects for TWFE and our simplified Sun--Abraham and Callaway--Sant'Anna implementations ($\rho=0.3$). The ordering shown is an artifact of measuring against $\delta$ rather than the cohort-share ATT; see text.
(c) Exogenous attrition (30\% missing unit-periods) further erodes power at moderate $N$ ($\rho=0.3$). Dashed lines mark the 80\% target; 1{,}000 replications per
cell.}
\label{fig:fig1and2}
\end{figure*}
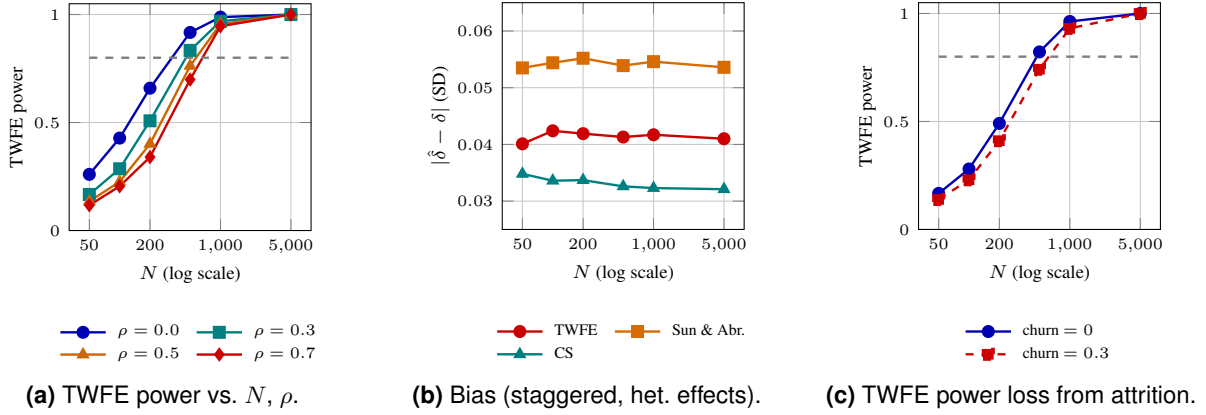

Two features of panel data drive a wedge between the power classical formulas
predict and the power TWFE delivers: within-unit serial correlation, and the
interaction between staggered timing and heterogeneous treatment effects.
Panels (a) and (b) of Figure~\ref{fig:fig1and2} summarize both.

Panel (a) reports TWFE power over $N$ and residual autocorrelation $\rho$,
holding \texttt{effect\_size}~$=0.2$~SD, a balanced panel, non-staggered
timing, and a 50\% pre-period. At $\rho=0$ --- the i.i.d.\ benchmark
underlying most closed-form power calculators --- $N=500$ reaches 92\% power,
consistent with the analytical formula. Raising $\rho$ to 0.5, a value
representative of weekly platform outcome series, reduces power at the same
$N$ to 76\%; at $\rho=0.7$ it falls to 70\%, and $N=200$ delivers only 34\%. The decay is steep
and non-linear, with the largest marginal losses between $\rho=0$ and
$\rho=0.5$.

Table~\ref{tab:nominal-gap} summarizes the nominal--achieved gap across
representative IS panel conditions. In the small-to-moderate-$N$ regime where
IS studies most commonly operate ($N \in [100, 1{,}000]$), the i.i.d.-versus-achieved gap ranges from $\approx6$~pp (at $N=1{,}000$) to $\approx33$~pp (at $N=500$ with attrition); the next paragraph decomposes how much of this is recoverable.

A natural objection is that this gap is measured against a \emph{naive} i.i.d.\ calculator, whereas competent applied work already clusters and anticipates serial correlation. We therefore decompose it against a stronger baseline: an AR(1)-aware, cluster-robust calculator, operationalized as the same simulation under the cell's $\rho$ but a balanced panel with correct unit clustering. This baseline is an \emph{oracle}---it uses the true $\rho$ and DGP---so the recoverability claims below are statements about that best-case upper bound. For balanced designs without attrition the baseline recovers essentially the entire gap: the serial-correlation loss is recoverable \emph{in principle} by modeling the dependence. What survives is attrition: at \texttt{effect\_size}~$=0.2$~SD and $\rho=0.5$, an AR(1)-aware design still overstates achievable power by roughly $9$~pp at $N=200$ and $11$~pp at $N=500$ once 30\% of unit-periods churn out (Table~\ref{tab:nominal-gap}; Monte Carlo SE $\approx2$~pp). The staggered-timing bias and pretest selection documented below are likewise invisible to any closed-form formula. The headline: serial correlation is recoverable; attrition, staggered timing, and pretesting are not.

How much of this recoverability survives when $\rho$ must be \emph{estimated}
rather than known? Table~\ref{tab:rho-sensitivity} quantifies the question
for the same design cells. With the five pre-periods these designs afford,
within-unit demeaning and the short estimation window jointly depress the
per-unit lag-1 correlation well below its true value: the naive plug-in
yields $\hat\rho\approx-0.12$ when the true $\rho$ is $0.3$, so a naive
plug-in planner behaves almost like the i.i.d.\ calculator it was meant to
replace, and the residual gap grows from $0$--$14$~pp (oracle) to
$7$--$35$~pp. Adding the first-order Nickell-type correction
$1/(T_{\text{pre}}-1)$ moves $\hat\rho$ to $\approx0.13$---about half of the
way back---and correspondingly recovers roughly half of the shortfall
($4$--$26$~pp); the remainder reflects small-sample bias in the within-unit
correlation itself, which the first-order term does not capture. Realizing the recovery on short pre-periods thus
requires bias-aware estimation of $\rho$ (or longer pre-periods); plugging a
raw demeaned-residual autocorrelation into a calculator---or into our lookup
tool's \texttt{autocorr} input---systematically overstates achievable power.

\begin{table}[t]
\centering
\small
\setlength{\tabcolsep}{4.5pt}
\begin{tabular}{@{}llrrrr@{}}
\toprule
$N$ & Condition & i.i.d. & AR(1) & Ach. & Resid. \\
\midrule
100     & $\rho{=}0.3$, no churn  & 0.29 & 0.20 & 0.20 & 0 \\
200     & $\rho{=}0.5$, churn 0.3 & 0.58 & 0.37 & 0.28 & 9 \\
500     & $\rho{=}0.5$, churn 0.3 & 0.92 & 0.70 & 0.59 & 11 \\
500     & $\rho{=}0.3$, no churn  & 0.94 & 0.80 & 0.80 & 0 \\
1{,}000 & $\rho{=}0.7$, no churn  & 1.00 & 0.94 & 0.94 & 0 \\
\bottomrule
\multicolumn{6}{@{}p{0.95\linewidth}@{}}{\footnotesize\emph{Resid.} $=$
AR(1)-aware planned power minus achieved power, in percentage points: the
part of the gap that survives a correctly specified AR(1)/cluster-robust
calculation.}\\
\end{tabular}
\caption{TWFE power across representative IS panel conditions
(\texttt{effect\_size}~$=0.2$~SD, $T=10$, 50\% pre-period, unit clustering)
under a naive i.i.d.\ calculator, an AR(1)-aware cluster-robust calculator
(same $\rho$, balanced panel), and achieved (simulation). Serial correlation
is recoverable (residual $\approx 0$ absent attrition); the surviving loss is
attrition. Monte Carlo SE $\approx1.5$--$2$~pp. Figure~\ref{fig:fig1and2}(a)
averages over $T\in\{6,10,16,24\}$, so its $\rho=0.5$ value (0.76) differs
from this table's $T=10$ AR(1) column (0.70).}
\label{tab:nominal-gap}
\end{table}

\begin{table}[t]
\centering
\small
\setlength{\tabcolsep}{1pt}
\begin{tabular}{@{}llrrrrr@{}}
\toprule
$N$ & Condition & Oracle & Plug-in & Corr. & Ach. & Resid. \\
\midrule
100     & $\rho{=}0.3$, no churn  & 0.23 & 0.33 & 0.28 & 0.23 & 0 / 10 / 5 \\
200     & $\rho{=}0.5$, churn 0.3 & 0.35 & 0.56 & 0.46 & 0.27 & 8 / 29 / 19 \\
500     & $\rho{=}0.5$, churn 0.3 & 0.72 & 0.93 & 0.84 & 0.58 & 14 / 35 / 26 \\
500     & $\rho{=}0.3$, no churn  & 0.79 & 0.93 & 0.89 & 0.80 & 0 / 13 / 9 \\
1{,}000 & $\rho{=}0.7$, no churn  & 0.92 & 1.00 & 0.97 & 0.93 & 0 / 7 / 4 \\
\bottomrule
\multicolumn{7}{@{}p{0.95\linewidth}@{}}{\footnotesize\emph{Resid.} $=$
planned minus achieved power (pp) under each baseline, reported as
oracle / plug-in / corrected.}\\
\end{tabular}
\caption{Planned power when $\rho$ is estimated from five pre-periods (500
pilot draws per cell; same cells as Table~\ref{tab:nominal-gap}; Monte Carlo
SE $\approx2$~pp). \emph{Oracle} plans at the true $\rho$; \emph{Plug-in} at
the raw demeaned-residual $\hat\rho$; \emph{Corr.} adds the first-order $1/(T_{\text{pre}}-1)$ demeaning-bias correction. Oracle and Ach.\ columns are re-simulated for this exercise and differ from Table~\ref{tab:nominal-gap} by Monte Carlo error.}
\label{tab:rho-sensitivity}
\end{table}

Panel (b) isolates the cost of staggered adoption with cohort-heterogeneous
effects (early cohorts receive $2\delta$, later cohorts $\delta$). Measured
against the base effect $\delta$, TWFE exhibits a persistent $\approx 0.042$
SD deviation that does not attenuate with $N$. Only part of this is the
\citet{goodmanbacon2021} distortion from negatively weighted ``forbidden''
comparisons; the remainder is a benchmark mismatch that the
reference-implementation check below isolates. Measured against the same base
effect, our implementation of \citet{callaway2021} shows a smaller deviation
($\approx0.033$ SD) and our implementation of \citet{sun2021} a larger one
($\approx0.054$ SD), stable across sample sizes (Monte Carlo SE
$\approx0.004$ SD).

We caution strongly against reading this ordering as an estimator ranking;
two robustness checks show it is an artifact. First, re-analyzing the full
grid, the apparent SA excess declines monotonically with pre-period length
(from $0.080$ SD at two pre-periods to $0.031$ SD at twelve) but persists
everywhere---pointing to something systematic. Second, we validated our
simplified implementations on a ten-cell subset (200 replications per cell)
against \texttt{fixest::sunab} (via its \texttt{pyfixest} port) and the
\texttt{differences} implementation of Callaway--Sant'Anna. Under this DGP
the two reference estimators are \emph{numerically identical} and
approximately unbiased for the realized cohort-share ATT---which under
heterogeneous effects is $0.23$--$0.30$~SD, not the base $\delta=0.2$.
Against that estimand, TWFE shows a modest $\approx{-}0.02$~SD attenuation
(the Goodman--Bacon distortion proper), simplified SA tracks the reference,
and simplified CS \emph{understates} the ATT by $\approx0.03$~SD via its
equal-weight aggregation. The panel-(b) ordering thus stacks a benchmark
mismatch (deviations from $\delta$ rather than the estimand) on simplified
aggregation weights: with reference implementations and the correct
estimand, the cohort-aware estimators are unbiased and effectively tied,
while TWFE retains a real but modest distortion. We retain the simplified
implementations only for the power and coverage columns of the main grid,
where they track the references at short panels (per-replication correlations $0.70$--$0.83$ at $T\le10$, falling to $\approx0.47$ in the two $T=24$ validation cells; $T=16$ was not validated).

\subsection{Panel Attrition and Effective Sample Size}
\label{sec:attrition}

Panel (c) of Figure~\ref{fig:fig1and2} isolates the effect of exogenous attrition on TWFE power at $\rho=0.3$ (\texttt{effect\_size}~$=0.2$~SD, non-staggered, averaged over $T$ and pre-period share). Moving
from \texttt{churn\_rate}$=0$ to 0.3 --- 30\% of unit-period observations
missing, drawn exogenously from both arms --- reduces power from 0.822 to
0.742 at $N=500$, and from 0.491 to 0.411 at $N=200$. In Table~\ref{tab:nominal-gap}'s $\rho = 0.5$, $T=10$ cells the contrast costs $9$--$11$~pp. The loss exceeds what a
naive ``effective sample size'' correction predicts: surviving observations
remain within-unit correlated, so usable signal contracts faster than the
observation count. The penalty is most pronounced at the sample sizes IS
studies most commonly operate in (roughly $N \in [200, 1000]$) and shrinks as
power saturates.

The main grid draws attrition identically from both arms, and the
$8$--$11$~pp penalty is specific to that mechanism. Because the more common
real-world failure mode is attrition that differs by treatment
status---adopters who churn after a disappointing experience, controls who
exit a comparison platform---we ran a dedicated 48-cell sub-grid
($N\in\{200,500\}$, $\rho\in\{0.3,0.5\}$, staggered and non-staggered,
1{,}000 replications per cell) varying the mechanism while holding total
attrition near 30\%. Two findings qualify the headline. First, attrition
that differs by treatment status but remains \emph{exogenous in timing}
(40\% of treated units versus 20\% of controls, or the mirror image) stays
essentially unbiased ($|\text{bias}| < 0.01$~SD) and costs only a few
additional percentage points of power ($0.72 \to 0.66$ at $N=500$,
$\rho=0.3$): it remains a power problem, and the lookup tool's single
\texttt{churn\_rate} input approximates it serviceably. Second,
\emph{outcome-dependent} attrition---elevated exit hazard after adverse
outcome shocks---changes the nature of the problem. Symmetric across arms,
the selection cancels in the DiD contrast; concentrated in the treated arm,
the estimate acquires an attenuation bias of $\approx{-}0.05$~SD, coverage
drops to $0.92$--$0.94$, and power falls to $0.44$ at $N=500$, $\rho=0.3$.
No power calculation repairs this case: it is a bias problem requiring
design or estimation remedies (tracking exiters, selection modeling,
bounds); the cohort-specific \citet{callaway2021} estimator is largely
robust to it in our staggered cells because its short cohort-window
comparisons limit exposure to post-exit selection. Attrition claims
elsewhere in the paper---including the abstract's $8$--$11$~pp
figure---refer to the exogenous case; outcome-dependent attrition is
strictly worse and differently shaped.

\subsection{Parallel-Trends Pretesting and Conditional Inference}

\begin{figure*}[t]
\centering
\begin{subfigure}[t]{0.24\linewidth}
\centering
\begin{tikzpicture}
\begin{axis}[compact, scale only axis, width=2.5cm, height=2.8cm,
  xlabel={Violation (SD/period)}, ylabel={Rejection rate},
  xtick={0,0.10,0.20,0.30}, xticklabel style={font=\tiny,/pgf/number format/.cd,fixed,precision=2},
  yticklabel style={font=\tiny},
  label style={font=\scriptsize},
  ymin=0, ymax=1.05,
  xlabel style={yshift=2pt},
  ylabel style={yshift=-4pt},
  legend style={at={(0.5,-0.42)},anchor=north,font=\tiny,
    column sep=1pt, /tikz/every even column/.append style={column sep=3pt}},
  legend columns=2]
\addplot[color=blue!70!black, mark=*]        coordinates {(0.00,0.034)(0.05,0.077)(0.10,0.276)(0.20,0.878)(0.30,0.999)}; \addlegendentry{$N=100$}
\addplot[color=orange!85!black, mark=square*] coordinates {(0.00,0.032)(0.05,0.328)(0.10,0.939)(0.20,1.000)(0.30,1.000)}; \addlegendentry{$N=500$}
\addplot[color=red!80!black, mark=triangle*] coordinates {(0.00,0.028)(0.05,0.633)(0.10,1.000)(0.20,1.000)(0.30,1.000)}; \addlegendentry{$N{=}1{,}000$}
\addplot[dotted, gray] coordinates {(0,0.05)(0.30,0.05)};
\end{axis}
\end{tikzpicture}
\caption{Pretrend rejection.}
\label{fig:fig3}
\end{subfigure}\hfill
\begin{subfigure}[t]{0.24\linewidth}
\centering
\begin{tikzpicture}
\begin{axis}[compact, scale only axis, width=2.5cm, height=2.8cm,
  xlabel={Violation (SD/period)}, ylabel={Rej.\ of $H_0{:}\beta{=}0$},
  xtick={0,0.10,0.20,0.30}, xticklabel style={font=\tiny,/pgf/number format/.cd,fixed,precision=2},
  yticklabel style={font=\tiny},
  label style={font=\scriptsize},
  ymin=0, ymax=1.05,
  xlabel style={yshift=2pt},
  ylabel style={yshift=-4pt},
  legend style={at={(0.5,-0.42)},anchor=north,font=\tiny,
    column sep=1pt, /tikz/every even column/.append style={column sep=3pt}},
  legend columns=1]
\addplot[color=blue!70!black, mark=*]     coordinates {(0.00,0.037)(0.05,0.351)(0.10,0.717)(0.20,0.936)(0.30,0.999)}; \addlegendentry{Unconditional}
\addplot[color=red!80!black,  mark=square*, dashed] coordinates {(0.00,0.028)(0.05,0.204)(0.10,0.249)(0.20,0.382)}; \addlegendentry{Pretest-pass}
\addplot[dotted, gray] coordinates {(0,0.05)(0.30,0.05)};
\end{axis}
\end{tikzpicture}
\caption{Cond.\ Type~I error.}
\label{fig:fig3a}
\end{subfigure}\hfill
\begin{subfigure}[t]{0.24\linewidth}
\centering
\begin{tikzpicture}
\begin{axis}[compact, scale only axis, width=2.5cm, height=2.8cm,
  xmode=log, log ticks with fixed point,
  xlabel={$N$ (log scale)}, ylabel={CI coverage},
  xtick={50,200,1000,5000},
  xticklabel style={font=\tiny},
  yticklabel style={font=\tiny},
  label style={font=\scriptsize},
  ymin=0.945, ymax=0.968,
  xlabel style={yshift=2pt},
  ylabel style={yshift=-4pt},
  legend style={at={(0.5,-0.42)},anchor=north,font=\tiny,
    column sep=1pt, /tikz/every even column/.append style={column sep=3pt}},
  legend columns=2]
\addplot[color=blue!70!black, mark=*]          coordinates {(50,0.9546)(100,0.9587)(200,0.9571)(500,0.9619)(1000,0.9596)(5000,0.9622)}; \addlegendentry{$\rho=0.0$}
\addplot[color=teal,           mark=square*]    coordinates {(50,0.9538)(100,0.9578)(200,0.9600)(500,0.9581)(1000,0.9609)(5000,0.9588)}; \addlegendentry{$\rho=0.3$}
\addplot[color=orange!85!black,mark=triangle*]  coordinates {(50,0.9521)(100,0.9571)(200,0.9601)(500,0.9578)(1000,0.9618)(5000,0.9602)}; \addlegendentry{$\rho=0.5$}
\addplot[color=red!80!black,   mark=diamond*]   coordinates {(50,0.9515)(100,0.9532)(200,0.9544)(500,0.9628)(1000,0.9603)(5000,0.9610)}; \addlegendentry{$\rho=0.7$}
\addplot[dashed, gray] coordinates {(50,0.95) (5000,0.95)};
\end{axis}
\end{tikzpicture}
\caption{TWFE 95\% CI coverage.}
\label{fig:fig8a}
\end{subfigure}\hfill
\begin{subfigure}[t]{0.24\linewidth}
\centering
\begin{tikzpicture}
\begin{axis}[compact, scale only axis, width=2.5cm, height=2.8cm,
  xmode=log, log ticks with fixed point,
  xlabel={$N$ (log scale)}, ylabel={CI coverage},
  xtick={50,200,1000,5000},
  xticklabel style={font=\tiny},
  yticklabel style={font=\tiny},
  label style={font=\scriptsize},
  ymin=0.945, ymax=0.968,
  xlabel style={yshift=2pt},
  ylabel style={yshift=-4pt},
  legend style={at={(0.5,-0.42)},anchor=north,font=\tiny,
    column sep=1pt, /tikz/every even column/.append style={column sep=3pt}},
  legend columns=2]
\addplot[color=blue!70!black, mark=*]          coordinates {(50,0.9554)(100,0.9593)(200,0.9572)(500,0.9620)(1000,0.9597)(5000,0.9622)}; \addlegendentry{$\rho=0.0$}
\addplot[color=teal,           mark=square*]    coordinates {(50,0.9550)(100,0.9582)(200,0.9603)(500,0.9582)(1000,0.9609)(5000,0.9588)}; \addlegendentry{$\rho=0.3$}
\addplot[color=orange!85!black,mark=triangle*]  coordinates {(50,0.9532)(100,0.9578)(200,0.9602)(500,0.9578)(1000,0.9618)(5000,0.9602)}; \addlegendentry{$\rho=0.5$}
\addplot[color=red!80!black,   mark=diamond*]   coordinates {(50,0.9528)(100,0.9538)(200,0.9546)(500,0.9630)(1000,0.9604)(5000,0.9610)}; \addlegendentry{$\rho=0.7$}
\addplot[dashed, gray] coordinates {(50,0.95) (5000,0.95)};
\end{axis}
\end{tikzpicture}
\caption{Sun--Abraham 95\% CI coverage.}
\label{fig:fig8b}
\end{subfigure}
\caption{Pretrend behavior and inferential calibration. (a) The pretrend
test has approximately nominal size and gains power with $N$.
(b) Conditioning on pretest passage inflates Type~I error ($\beta=0$ DGP;
the conditional curve ends at 0.20 because the pretest rejects everywhere at
larger violations). (c)--(d) 95\% CI coverage for TWFE and Sun--Abraham
(non-staggered, homogeneous, unit clustering).}
\label{fig:fig3both}
\end{figure*}
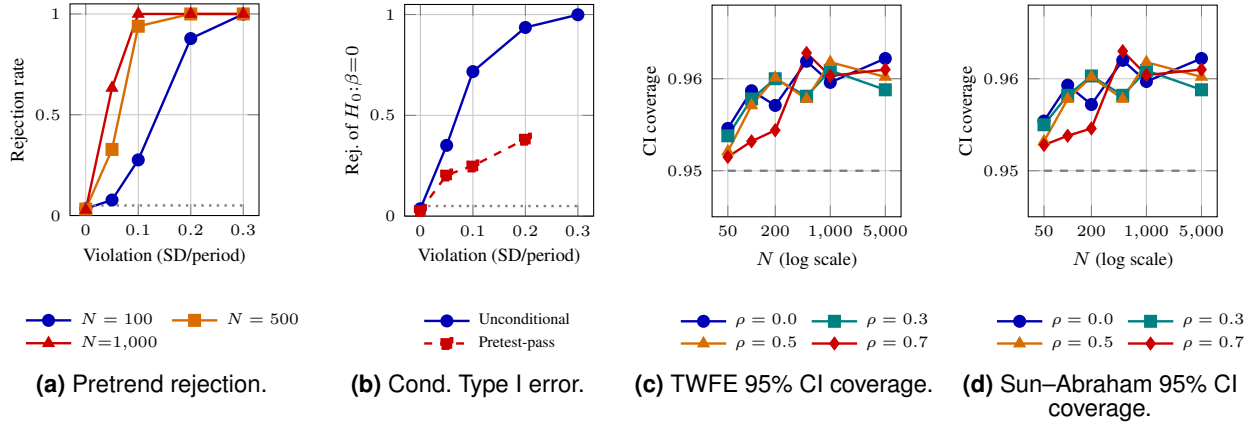

We next examine pretesting for parallel trends. Figure~\ref{fig:fig3both}(a)
reports the rejection rate of the joint Wald test on pre-period event-time
coefficients: under a true null it rejects at 0.028--0.037 across all sample
sizes (approximately correct size), and its power rises monotonically in the
violation magnitude and in $N$ (at $N=1{,}000$, violations of 0.05 SD/period
are detected in 63\% of replications).

The substantive concern, as \citet{roth2022} emphasizes, is the inferential
consequence of \emph{conditioning} on having passed the pretest.
Figure~\ref{fig:fig3both}(b) uses a DGP with a \emph{true null} treatment
effect ($\beta=0$) but non-zero pre-trend violations, so any rejection of
$H_0:\beta=0$ is a false positive. At a violation of 0.10 SD/period,
unconditional rejection is 0.717, while among replications that pass the
pretest, conditional rejection remains 0.249---nearly five times the nominal
0.05 level. The mechanism is selection: pretest-passing replications are
those in which sampling noise attenuates the estimated pre-trend, and the
same noise propagates into a biased post-treatment estimate. A binary
``pretest-gate'' rule therefore confers far less Type~I error protection than
its nominal size suggests, and the gap widens for the small violations most
likely to escape detection.

\subsection{Confidence-Interval Coverage}
\label{sec:coverage}

Panels (c) and (d) of Figure~\ref{fig:fig3both} report empirical 95\% CI
coverage for TWFE and Sun--Abraham across $N$ and $\rho$ under the
non-staggered, homogeneous benchmark. Coverage remains near nominal
(0.951--0.963) with no systematic drift, indicating that the power losses
above reflect effective-sample-size contraction rather than under-coverage.
For Callaway--Sant'Anna, whose simplified across-cohort standard error is
undefined in this single-cohort benchmark, a unit-level bootstrap yields
coverage of $0.91$--$0.97$ (mean $\approx0.94$) across $N\in\{200,500\}$ and
$\rho\in\{0.3,0.5\}$---near-nominal and comparable to the other estimators.

The picture is different in the \emph{staggered} cells, where the simplified
C\&S analytic interval---the across-cohort dispersion of per-cohort
ATTs---undercovers systematically. Across all $2{,}304$ staggered
homogeneous-effects cells its coverage averages $0.906$, never exceeds
$0.955$, and one third of cells fall below $0.90$. The undercoverage is flat
in $N$ and improves only with panel length ($0.856$ at $T=6$ rising to
$0.934$ at $T=24$): the signature of a standard error estimated from a
handful of cohort-level ATTs, whose count grows with $T$ but not $N$. The
supplement's worked example ($0.876$, a staggered heterogeneous cell with
attrition) is part of this pattern, not an outlier. By comparison, TWFE
coverage under unit clustering in the same cells is nominal ($0.957$) and
Sun--Abraham close to it ($0.937$); a separate tail arises under
\emph{two-way} clustering, where the truncated CGM variance estimator
undercovers (mean $0.865$, worsening with $T$)---a distinct issue we flag
but do not pursue. Practically: treat the simplified C\&S analytic interval
in staggered designs as anti-conservative by roughly $4$--$9$ coverage
points, and prefer the reference multiplier-bootstrap inference of the
\texttt{did} package whenever its intervals matter for a conclusion.

\subsection{Supplementary DiD Robustness Modules}

Beyond the core grid, a treatment-process module allows non-absorbing and continuous-dose exposure. Binary TWFE degrades under continuous treatment (bias $\approx 0.101$, RMSE $\approx 0.143$, power $\approx 0.491$) while absorbing and non-absorbing designs stay near-unbiased with coverage $\approx 0.95$--$0.96$: treatment-process misspecification can be first-order. We restrict supplementary estimators to specifications we can implement faithfully, leaving imputation, matrix-completion, and causal-ML comparisons to their reference implementations in follow-up work.

\subsection{Empirical Anchor: Classic DiD Design}

To anchor the simulation evidence in observed data, we replicate a classic
open DiD panel design using the firm-level job-training dataset
(54 firms, 3 years; grant adoption in 1988--1989; \cite{wooldridge2010}).
The estimated TWFE effect is $-0.279$ (SE $0.198$). Treating this magnitude
as an \emph{ex ante} design anchor---a posited target effect rather than
observed power, which is an uninformative transform of the $p$-value
\citep{hoenigheisey2001}---the design-based i.i.d.\ power for this panel is
$0.291$, the calculation most frequently reported in applied IS design
sections. Mapping the design to the nearest cells in the simulation grid
(staggered adoption, unit clustering, nearest $N$, $T$, pre-period share,
autocorrelation, and effect size) yields materially lower achieved power:
$0.067$ for TWFE, $0.097$ for Sun--Abraham, and $0.134$ for
Callaway--Sant'Anna. Because the anchor is mapped to the \emph{nearest} grid cell rather than
simulated at its exact parameters, part of this gap reflects cell
coarseness, and here that part cannot be tightly bounded: the panel's three
periods lie below the grid's minimum $T=6$, so the mapped cell is a coarse
approximation. For the IV anchor below, an exact-parameter replication
bounds the corresponding contribution at about one percentage point. The nominal-achieved gap for TWFE is about
22 percentage points, consistent with the main result that classical power
heuristics overstate inferential strength in short, persistent panels.

\section{IV Results}

IV identification rests on two conceptually independent requirements ---
relevance and exclusion --- often evaluated through the single lens of
first-stage diagnostics. Our IV grid separates them, varying 2SLS performance
over first-stage strength, sample size, and direct $Z \to Y$ violations of
exclusion, with the true structural effect at $\beta=0.1$ (panels (a)--(b) of
Figure~\ref{fig:iv}) or $\beta=0$ (panels (c)--(d)).

\subsection{Power as a Function of First-Stage Strength and Sample Size}

\begin{figure*}[t]
\centering
\begin{subfigure}[t]{0.24\linewidth}
\centering
\begin{tikzpicture}
\begin{axis}[compact, scale only axis, width=2.5cm, height=2.8cm,
  xmode=log, log ticks with fixed point,
  xlabel={First-stage $F$}, ylabel={IV power},
  xtick={5,10,50,100},
  xticklabel style={font=\tiny},
  yticklabel style={font=\tiny},
  label style={font=\scriptsize},
  ymin=0, ymax=0.25,
  xlabel style={yshift=2pt},
  ylabel style={yshift=-4pt},
  legend style={at={(0.5,-0.42)},anchor=north,font=\tiny,
    column sep=1pt, /tikz/every even column/.append style={column sep=3pt}},
  legend columns=2]
\addplot[color=blue!30!black, mark=*] coordinates {(5,0.088)(10,0.089)(20,0.099)(50,0.135)(100,0.155)}; \addlegendentry{$N{=}100$}
\addplot[color=blue!60!black, mark=square*] coordinates {(5,0.075)(10,0.080)(20,0.105)(50,0.131)(100,0.162)}; \addlegendentry{$N{=}300$}
\addplot[color=teal, mark=triangle*] coordinates {(5,0.071)(10,0.088)(20,0.089)(50,0.149)(100,0.154)}; \addlegendentry{$N{=}500$}
\addplot[color=orange!85!black, mark=diamond*] coordinates {(5,0.068)(10,0.086)(20,0.108)(50,0.135)(100,0.187)}; \addlegendentry{$N{=}1{,}000$}
\addplot[color=red!70!black, mark=pentagon*] coordinates {(5,0.072)(10,0.099)(20,0.087)(50,0.124)(100,0.191)}; \addlegendentry{$N{=}3{,}000$}
\addplot[color=red!30!black, mark=otimes*] coordinates {(5,0.068)(10,0.100)(20,0.105)(50,0.141)(100,0.183)}; \addlegendentry{$N{=}10$k}

\end{axis}
\end{tikzpicture}
\caption{IV power ($\beta{=}0.1$).}
\label{fig:iv-a}
\end{subfigure}\hfill
\begin{subfigure}[t]{0.24\linewidth}
\centering
\begin{tikzpicture}
\begin{axis}[compact, scale only axis, width=2.5cm, height=2.8cm,
  xmode=log, log ticks with fixed point,
  xlabel={$N$ (log scale)}, ylabel={$|\hat\beta-\beta|$},
  xtick={100,500,3000,10000},
  xticklabel style={font=\tiny},
  yticklabel style={font=\tiny},
  label style={font=\scriptsize},
  ymin=0, ymax=0.1,
  xlabel style={yshift=2pt},
  ylabel style={yshift=-4pt},
  legend style={at={(0.5,-0.42)},anchor=north,font=\tiny,
    column sep=1pt, /tikz/every even column/.append style={column sep=3pt}},
  legend columns=2]
\addplot[color=red!70!black,   mark=*]        coordinates {(100,0.022)(300,0.049)(500,0.022)(1000,0.084)(3000,0.010)(10000,0.031)}; \addlegendentry{$F{=}5$}
\addplot[color=orange!85!black,mark=square*]  coordinates {(100,0.001)(300,0.006)(500,0.013)(1000,0.082)(3000,0.013)(10000,0.003)}; \addlegendentry{$F{=}10$}
\addplot[color=teal,           mark=triangle*] coordinates {(100,0.009)(300,0.014)(500,0.004)(1000,0.003)(3000,0.005)(10000,0.001)}; \addlegendentry{$F{=}20$}
\addplot[color=blue!60!black,  mark=diamond*]  coordinates {(100,0.008)(300,0.001)(500,0.001)(1000,0.002)(3000,0.004)(10000,0.003)}; \addlegendentry{$F{=}50$}
\addplot[color=blue!30!black,  mark=pentagon*] coordinates {(100,0.003)(300,0.003)(500,0.002)(1000,0.000)(3000,0.003)(10000,0.001)}; \addlegendentry{$F{=}100$}
\end{axis}
\end{tikzpicture}
\caption{Weak-IV bias.}
\label{fig:iv-b}
\end{subfigure}\hfill
\begin{subfigure}[t]{0.24\linewidth}
\centering
\begin{tikzpicture}
\begin{axis}[compact, scale only axis, width=2.5cm, height=2.8cm,
  xmode=log, log ticks with fixed point,
  xlabel={$N$ (log scale)}, ylabel={$\hat\beta-\beta$},
  xtick={100,500,3000,10000},
  xticklabel style={font=\tiny},
  yticklabel style={font=\tiny},
  label style={font=\scriptsize},
  ymin=-0.05, ymax=1.7,
  xlabel style={yshift=2pt},
  ylabel style={yshift=-4pt},
  legend style={at={(0.5,-0.42)},anchor=north,font=\tiny,
    column sep=1pt, /tikz/every even column/.append style={column sep=3pt}},
  legend columns=2]
\addplot[color=blue!60!black, mark=*]        coordinates {(100,-0.003)(300,-0.003)(500,-0.002)(1000,0.000)(3000,0.003)(10000,0.001)}; \addlegendentry{$\gamma{=}0.00$}
\addplot[color=teal,          mark=square*]  coordinates {(100,0.052)(300,0.082)(500,0.097)(1000,0.128)(3000,0.221)(10000,0.398)}; \addlegendentry{$\gamma{=}0.05$}
\addplot[color=orange!85!black,mark=triangle*] coordinates {(100,0.112)(300,0.161)(500,0.195)(1000,0.265)(3000,0.441)(10000,0.794)}; \addlegendentry{$\gamma{=}0.10$}
\addplot[color=red!70!black,  mark=diamond*] coordinates {(100,0.227)(300,0.315)(500,0.384)(1000,0.520)(3000,0.878)(10000,1.588)}; \addlegendentry{$\gamma{=}0.20$}
\addlegendimage{empty legend}\addlegendentry{\vphantom{$F$}}
\addlegendimage{empty legend}\addlegendentry{\vphantom{$F$}}
\end{axis}
\end{tikzpicture}
\caption{Exclusion bias ($F{=}100$).}
\label{fig:iv-c}
\end{subfigure}\hfill
\begin{subfigure}[t]{0.24\linewidth}
\centering
\begin{tikzpicture}
\begin{axis}[compact, scale only axis, width=2.5cm, height=2.8cm,
  xmode=log, log ticks with fixed point,
  xlabel={First-stage $F$}, ylabel={Type~I error},
  xtick={5,10,50,100},
  xticklabel style={font=\tiny},
  yticklabel style={font=\tiny},
  label style={font=\scriptsize},
  ymin=0, ymax=0.08,
  xlabel style={yshift=2pt},
  ylabel style={yshift=-4pt},
  legend style={at={(0.5,-0.42)},anchor=north,font=\tiny,
    column sep=1pt, /tikz/every even column/.append style={column sep=3pt}},
  legend columns=2]
\addplot[color=blue!70!black, mark=*]          coordinates {(5,0.016)(10,0.026)(20,0.036)(50,0.044)(100,0.053)}; \addlegendentry{$\rho_{vu}{=}0.3$}
\addplot[color=orange!85!black, mark=square*]  coordinates {(5,0.042)(10,0.044)(20,0.044)(50,0.043)(100,0.047)}; \addlegendentry{$\rho_{vu}{=}0.5$}
\addplot[color=red!80!black, mark=triangle*]   coordinates {(5,0.069)(10,0.069)(20,0.053)(50,0.056)(100,0.053)}; \addlegendentry{$\rho_{vu}{=}0.7$}
\addlegendimage{empty legend}\addlegendentry{\vphantom{$F$}}
\addlegendimage{empty legend}\addlegendentry{\vphantom{$F$}}
\addlegendimage{empty legend}\addlegendentry{\vphantom{$F$}}
\addplot[dashed, gray] coordinates {(5,0.05)(100,0.05)};
\end{axis}
\end{tikzpicture}
\caption{IV Type~I error ($\beta{=}0$).}
\label{fig:iv-d}
\end{subfigure}
\caption{IV diagnostics across first-stage strength, sample size, and
assumption failures. (a)~Power never exceeds $\approx0.19$---far below the
0.80 target---even at $N=10{,}000$ and $F=100$ ($\beta=0.1$,
$\rho_{vu}=0.5$, single instrument). (b)~Weak-instrument bias attenuates
with $F$ and $N$; fluctuations at $F\in\{5,10\}$, $N=1{,}000$ are Monte
Carlo noise. (c)~With $F=100$ but a direct $Z\to Y$ channel $\gamma$, bias
scales \emph{with} $N$ (true $\beta=0$, $\rho_{vu}=0.5$).
(d)~Heteroskedasticity-robust 2SLS holds size near 0.05 (dashed) even at
$F=5$; distortion rises only at $\rho_{vu}=0.7$.}
\label{fig:iv}
\end{figure*}
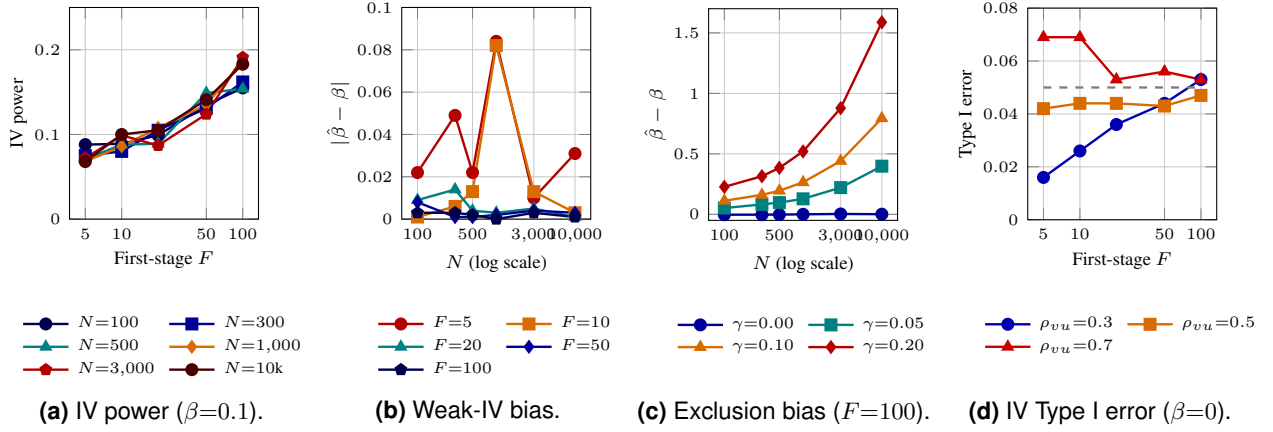

Panel~(a) of Figure~\ref{fig:iv} plots IV power over the target first-stage $F$ and
$N$, conditional on a true structural effect of $\beta=0.1$ (a ``small''
effect by the Table~\ref{tab:is-anchors} anchors, and the range most common
in IS platform and recommender studies). For this effect size, 80\% power is
infeasible within our grid: power never exceeds $\approx0.19$, even at
$F=100$ and $N=10{,}000$. At $F=10$ --- the canonical \citet{stockyogo2005}
reference point --- power is $\approx0.08$--$0.10$ regardless of $N$, and in
the $F=10$--$20$ range typical of IS instruments (geographic variation, peer
exposure) even $N=1{,}000$ delivers power below 0.11. The ceiling reflects
the fixed-$F$ normalization: holding $F\approx100$ pins the concentration
parameter near 100 regardless of $N$, so $\mathrm{SE}(\hat\beta)$ stays
$\approx0.10$ and $|\hat\beta|/\mathrm{SE}\approx1$ for $\beta=0.1$, far
short of conventional power. This is a statement about fixed $F$, not data
collection: with a \emph{fixed} instrument (constant first-stage
coefficient), $F$ grows with $N$, $\mathrm{SE}(\hat\beta)$ falls as
$N^{-1/2}$, and power rises in the usual way. The grid holds $F$ fixed
precisely to isolate its role, so the lever it identifies is instrument
strength, not sample size---and the $F>10$ convention, while informative
about size, is an inadequate proxy for power when the target effect is
small.

\subsection{Two Failure Modes: Weak Instruments versus Exclusion Violations}

The two canonical IV assumption failures --- a weak first stage and a
non-zero direct $Z \to Y$ channel --- produce bias signatures that, under our
fixed-$F$ normalization, scale with $N$ in opposite directions. Panel~(b) of
Figure~\ref{fig:iv} plots the absolute finite-sample bias of 2SLS as $F$
ranges from 5 to 100. At $F \ge 50$ bias is in the $10^{-3}$ range across
all sample sizes; at $F=5$ it is visible but attenuates as $N$ grows,
consistent with weak-instrument bias being $O(N^{-1})$ conditional on
identification \citep{angristpischke2009}.

Panel~(c) of Figure~\ref{fig:iv} fixes $F$ at 100 and varies the direct-effect
parameter $\gamma$ governing the $Z \to Y$ channel outside the structural
equation. Because the design holds the target $F$ fixed,
$\pi=\sqrt{F/(N+F)}\propto N^{-1/2}$, and the just-identified 2SLS plim bias
from a direct channel is $\gamma/\pi$, so it grows as $\sqrt{N}$: for
$\gamma=0.05$ it rises from 0.052 at $N=100$ to 0.398 at $N=10{,}000$
($7.7\times$ observed versus $7.1\times$ predicted). Under a fixed
instrument coefficient instead, the plim bias $\gamma/\pi$ is constant in
$N$ while sampling variance shrinks---the estimator becomes confidently
wrong. The message shared across both normalizations is narrow but firm:
sample size does not cure exclusion bias. The asymmetry with panel~(b) has
direct diagnostic implications: the standard toolkit (first-stage $F$,
Hansen $J$) cannot separate the two regimes with a single-instrument design,
so over-identification tests provide limited leverage in exactly the
settings where the bias grows fastest.

\subsection{Size Properties of Robust 2SLS Under Weak Instruments}

Panel~(d) of Figure~\ref{fig:iv} sets $\beta=0$ and reports the Type~I error
of heteroskedasticity-robust 2SLS. Even at $F=5$, size never exceeds 0.069,
and it lies above 0.05 only for strong endogeneity ($\rho_{vu}=0.7$); for
$F \ge 20$ it is indistinguishable from nominal. Size distortion is thus more
limited than classical \citet{stockyogo2005} warnings imply, consistent with
\citet{angristpischke2009} and \citet{semadeni2014} on robust standard
errors (see \cite{londschien2025} for a recent overview). Weak
instruments in our DGP damage designs primarily through low power and
first-stage bias rather than inflated false-positive rates.

\subsection{Empirical Anchor: Classic IV Design}

We also replicate a standard open IV design using the Card returns-to-schooling
dataset (instrument: \texttt{nearc4}, $N=3{,}010$; \cite{card1995}). The
estimated 2SLS coefficient on schooling is $0.119$ (SE $0.053$), with
first-stage $F=14.49$. Treating $0.119$ as an \emph{ex ante} design anchor
\citep{hoenigheisey2001}, the design-based i.i.d.\ power is $0.614$. Mapping
this design to the nearest IV-grid cell ($N=3{,}000$, $F=20$,
$\rho_{vu}=0.3$, $\beta=0.1$) gives achieved power of $0.060$, a
nominal-achieved gap of roughly 55 percentage points. The gap is not an artifact of nearest-cell coarseness: a
targeted replication at Card's exact parameters ($\beta=0.119$, $F=14.49$,
$N=3{,}010$) shifts achieved power by under one percentage point, bounding
the cell-coarseness contribution for this anchor at about one point. The
remainder reflects a weak first stage interacting with a small true effect:
clearing conventional relevance heuristics does not guarantee adequate power
at Card's \emph{realized} $N$ and $F$. Because \texttt{nearc4} is a fixed
instrument, a larger-$N$ version of the same design would have
proportionally higher $F$ and materially higher power; the lesson concerns
instrument strength at the available sample, not the futility of collecting
more data.

\section{Discussion}

\subsection{Implications for Study Design}

We offer these as evidence-based guidance, not prescriptions.
(1)~Residual autocorrelation is a first-order but \emph{recoverable} input to
DiD power (Table~\ref{tab:nominal-gap}). Where pre-period data exist,
estimate $\rho$ from unit-demeaned residuals---with the bias correction of
Table~\ref{tab:rho-sensitivity} on short pre-periods---and propagate it
through a calibrated simulation; otherwise report a power curve over
$\rho\in[0,0.7]$.
(2)~What is \emph{not} recoverable---attrition and staggered timing---must be
built into the design directly, and is better read off the companion lookup
tool's nine-dimensional cell than 2D slices.
(3)~Estimator choice matters, and so do implementation fidelity and the
estimand: reference implementations of Sun--Abraham and Callaway--Sant'Anna
are approximately unbiased for the cohort-share ATT while TWFE attenuates it
by $\approx0.02$~SD, but orderings among \emph{simplified} reimplementations
can invert that conclusion (Section~\ref{sec:did-results}). Report TWFE
alongside a cohort-aware estimator \emph{from a reference package} and state
the estimand explicitly.
(4)~Binary pretesting inflates conditional Type~I error \citep{roth2022}; prefer
reporting the unconditional estimate with a sensitivity analysis
\citep{rambachanroth2023} over a pass/fail gate.
(5)~The $F>10$ convention speaks to size, not power (at $F=10$, $N=500$, IV
power $<0.11$); report achieved power explicitly.
(6)~Exclusion bias is not cured by sample size; defend exclusion on
substantive or design grounds, not via larger panels.
(7)~CI coverage is a useful second-order diagnostic: deviations from $0.95$
signal mis-specification such as inadequate clustering.
(8)~Advanced estimators (imputation, matrix completion, causal ML) rely on their reference inference machinery; validate coverage on a study-like DGP before trusting their default intervals.

\subsection{Limitations}
Our DGP uses AR(1) residuals, unit and time effects, and a simple trend; multi-way error dependence in the DGP (we vary only the clustering used for inference), unit-specific trends, and seasonality are out of scope,
so the irreducible gap we isolate is a conservative lower bound for messier
panels. The attrition sub-grid covers treatment-differential and
outcome-dependent exit, but not richer dynamic selection (anticipation,
selection on effect size). Our simplified Callaway--Sant'Anna
analytic SE undercovers across the staggered grid
(Section~\ref{sec:coverage}); porting the reference \texttt{did}
multiplier-bootstrap SE across the full grid is future work. We do not
evaluate parallel-trends sensitivity analyses \citep{rambachanroth2023}; the
main DiD grid uses \texttt{effect\_size}~$>0$ (DiD Type~I error is taken
from the pretest and IV modules); Monte Carlo precision is $1{,}000$
replications per cell; and, as noted at the point of each claim, the
AR(1)-aware baseline is an oracle (Table~\ref{tab:rho-sensitivity}
quantifies the estimated-$\rho$ case) and each empirical anchor is mapped to
the nearest grid cell (coarseness bounded for IV by the exact-parameter Card replication; unbounded for the DiD anchor, whose $T$ lies outside the grid).

\subsection{Next Steps: Toward IS Research Methodology for the 2030s}
The grid---an empirically anchored, validated simulation released as a
queryable artifact---is a prototype for how ex ante design evidence in IS
could be produced and consumed. Four next steps: richer failure modes
(informative missingness, dynamic selection, anticipation---mapping rather
than assuming the power-versus-bias boundary); estimator fidelity as
infrastructure, extending reference Sun--Abraham, Callaway--Sant'Anna
(multiplier-bootstrap inference), imputation \citep{borusyak2024,
gardner2022}, matrix-completion \citep{athey2021}, and causal-ML
\citep{chernozhukov2018, wagerathey2018} estimators across the full grid so
the community compares estimators on equal footing rather than through
per-paper reimplementations; design tooling that folds estimated-parameter
uncertainty and parallel-trends sensitivity \citep{rambachanroth2023} into
the lookup tool for registration-ready workflows; and a living, versioned
benchmark open to community-contributed cells, as shared benchmarks did for
machine learning. As IS panels grow larger and messier, we expect
empirically calibrated simulation evidence to displace closed-form power
formulas as the default justification standard.

\subsection{Reproducibility and Computational Cost}
The pipeline is implemented in Python with per-cell hash-derived seeds, so any cell is independently reproducible. The full grids ($\sim$9.8 million datasets) required roughly four days of wall-clock time on a 14-core consumer laptop---no cluster access was used or needed. This cost matters mainly for full replication: a single design cell reproduces in minutes and a targeted sub-grid within an hour on commodity hardware, and the lookup tool removes the need to re-simulate at all for designs inside the released grid. 

\section*{Acknowledgment}
LLMs were used strictly for language polishing (grammar and readability); we reviewed and verified all content and assume full responsibility for its accuracy and integrity.

\printbibliography

\end{document}